\documentclass[final]{svjour2}
\usepackage{graphicx}
\usepackage{rotating}
\usepackage{amssymb}
\usepackage{mathptmx}
\usepackage[numbers]{natbib}
\makeatletter
\journalname{Journal of Low Temperature Physics}
\bibpunct{}{}{,}{s}{}{,}

\begin{document}

\newcommand{\hdblarrow}{H\makebox[0.9ex][l]{$\downdownarrows$}-}
\title{Magnetization curve of the kagome-strip-lattice antiferromagnet}

\author{T. Shimokawa$^1$ \and H. Nakano$^1$}

\institute{1:University of Hyogo, Hyogo, Japan\\
%Tel.:\\ Fax:\\
\email{t.shimokaw@gmail.com}
}

\date{10.06.2012}

\maketitle


\begin{abstract}

%We have studied the magnetization curve of the Heisenberg model on the quasi-one-dimensional kagome-strip lattice that shares the same lattice structure 
%in the inner part with the two-dimensional kagome lattice.
%Our numerical calculation by using the density matrix renormalization group method reveal that the system has the magnetization plateaus at $n$/7 of the full magnetization. 
%We find the presence of the magnetic plateaus of $n=1, 2, 3, 4, 5$ and 6 in the $S=1/2$ case.
%In the $S=1$ case, on the other hand, we have clearly found the magnetic plateaus of $n=1, 3, 5$ and 6.
%In particular, we confirm the translational symmetry breaking from our calculations of the local magnetizations on the plateaus of $n=2,4$ and 6 as predicted by the Oshikawa-Yamanaka-Affleck condition. 
%We have also found the occurrence of the macroscopic jump in these magnetization curves near the saturation field irrespective of the spin amplitude as well as the original kagome model.

We study the magnetization curve of the Heisenberg model on
the quasi-one-dimensional kagome-strip lattice that shares the same lattice structure
in the inner part with the two-dimensional kagome lattice. 
Our numerical calculations based on the density matrix renormalization group method 
reveal that the system shows several magnetization plateaus 
between zero magnetization and the saturated one; 
we find the presence of the magnetic plateaus with the $n/7$ height of the saturation 
for $n=$1,2,3,4,5 and 6 in the $S=$1/2 case, 
whereas we detect only the magnetic plateaus of $n=$1,3,5 and 6 in the $S=$1 case. 
In the cases of $n=$2,4 and 6 for the $S=$1/2 system, the Oshikawa-Yamanaka-Affleck 
condition suggests the occurrence of the translational symmetry breaking (TSB). 
We numerically confirm this non-trivial TSB in our results of local magnetizations. 
We have also found that the macroscopic jump appears near the saturation field 
irrespective of the spin amplitude as well as the two-dimensional kagome model.

%75.10.Jm	Quantized spin models, including quantum spin frustration
%75.30.Kz	Magnetic phase boundaries (including classical and quantum magnetic transitions, metamagnetism, etc.) (for ferroelectric phase transitions, see 77.80.B-; for superconductivity phase diagrams, see 74.25.Dw)
%75.45.+j	Macroscopic quantum phenomena in magnetic systems

\end{abstract}

\section{Introduction}
\begin{figure}
\begin{center}
\includegraphics[%
  width=1\linewidth,
  keepaspectratio]{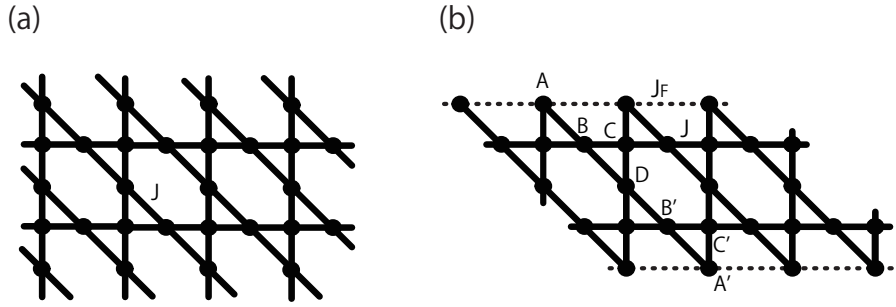}
\end{center}
\caption{(Color online) Structures of the lattices: the kagome lattice (a), the quasi-one-dimensional
kagome strip lattice (b).Antiferromagnetic bonds $J$ (bold straight line) 
and ferromagnetic bond $J_{\rm F}$ (dotted line). 
Sublattices in a unit cell of lattice (b) are represented 
by A, ${\rm A}^{\prime}$, B, ${\rm B}^{\prime}$, 
C, ${\rm C}^{\prime}$, and D.
 }
\label{fig1}
\end{figure}

The kagome-lattice antiferromagnet has been extensively studied from both of experimental and theoretical approaches 
in the high expectation that the effect of the large quantum fluctuation and the strong frustration produce the exotic states.\cite{KGM1, KGM2, KGM3, KGM4, KGM5, KGM6, KGM7, KGM8, KGM9, KGM10, KGM11}
From the viewpoint of the numerical studies, however, it is well known that 
the applications of the existing methods, the density matrix renormalization group (DMRG)\cite{DMRG1, DMRG2}, quantum monte carlo (QMC) and exact diagonalization (ED) methods, to the two-dimensional frustrated system such as the kagome antiferromagnet are difficult.
The implementation of the DMRG method is difficult in more than two dimensions although this method is very powerful for one- or quasi-one-dimensional systems under the open boundary condition and the QMC method comes across the so-called negative-sign problem for frustrated systems. 
The ED method has the strong limitation of available system sizes 
although this method does not suffer from the limitation of the dimensionality nor negative sign problem.
Due to these difficulties, there still remain many unresolved problems about the nontrivial natures of the kagome antiferromagnets.
In particular, very recent ED studies on the $S=1/2$ antiferromagnetic Heisenberg model on the kagome lattice depicted in Fig.~\ref{fig1}(a) clarified 
that there exists an anomalous behavior at the 1/3 height of the saturation in the magnetization process, the magnetization ramp \cite{KGM9, KGM10}
whose critical behavior is quite different from conventional magnetization plateau and magnetization cusp.
Therefore, the magnetization curve of the kagome antiferromagnet is attracting the most attention.

In this study, we investigate the magnetization curve of the Heisenberg model on the quasi-one-dimensional (Q1D) kagome strip lattice 
depicted in Fig.~\ref{fig1}(b) instead of the 2D lattice depicted in Fig.~\ref{fig1}(a). 
Note that the inner parts of the lattices in Fig.~\ref{fig1}(b) are common to a part of the 2D lattice in Fig.~\ref{fig1}(a). 
This model was originally introduced in ref. 14 for studying on the occurrence of the non-Lieb-Mattis ferrimagnetism \cite{Yoshikawa, Hida, strip1} 
in the ground-state of the $S=1/2$ antiferromagnetic Heisenberg model on the spatially anisotropic kagome lattice\cite{KGM12}.
However, the magnetization curve of this Q1D system has not been investigated in detail.
We will show our numerical results obtained by the DMRG calculations not only in the $S=1/2$ case but also $S=1$ case. 

\section{Model Hamiltonian}
The Hamiltonian of the present model is given by
\begin{eqnarray}
\label{Hamiltonian}
\mathcal{H} &=&
  J \sum_{i}  [ {\bf S}_{i,{\rm B}}\cdot {\bf S}_{i,{\rm C}} 
                     + {\bf S}_{i,{\rm C}}\cdot {\bf S}_{i,{\rm D}}
                     + {\bf S}_{i,{\rm C}}\cdot {\bf S}_{i+1,{\rm A}}  
                     + {\bf S}_{i,{\rm C}}\cdot {\bf S}_{i+1,{\rm B}}
 \nonumber \\
                   &+& {\bf S}_{i,{\rm C}^{\prime}}\cdot {\bf S}_{i,{\rm B}^{\prime}}
                     + {\bf S}_{i,{\rm C}^{\prime}}\cdot {\bf S}_{i,{\rm A}^{\prime}}
                     + {\bf S}_{i,{\rm C}^{\prime}}\cdot {\bf S}_{i+1,{\rm D}}
                     +{\bf S}_{i,{\rm C}^{\prime}}\cdot {\bf S}_{i+1,{\rm B}^{\prime}} 
\nonumber \\
&+&J                 {\bf S}_{i,{\rm A}}\cdot {\bf S}_{i,{\rm B}}
                         +{\bf S}_{i,{\rm B}}\cdot {\bf S}_{i,{\rm D}}
                         +{\bf S}_{i,{\rm D}}\cdot {\bf S}_{i,{\rm B}^{\prime}}
                         +{\bf S}_{i,{\rm B}^{\prime}}\cdot {\bf S}_{i,{\rm A}^{\prime}}]                          
\nonumber \\
&+&                     
J_{\rm F} \sum_{i} [{\bf S}_{i,{\rm A}}\cdot {\bf S}_{i+1,{\rm A}}
                     +{\bf S}_{i,{\rm A}^{\prime}}\cdot {\bf S}_{i+1, {\rm A}^{\prime}}]
\nonumber \\
&-& h \sum_{i} [S^{z}_{i,{\rm A}}+S^{z}_{i,{\rm A}^{\prime}}+S^{z}_{i,{\rm B}}+S^{z}_{i,{\rm B}^{\prime}}+S^{z}_{i,{\rm C}}+S^{z}_{i,{\rm C}^{\prime}}+S^{z}_{i,{\rm D}}],
\end{eqnarray}
where  ${\bf S}_{i, \xi}$ is an $S=1/2$ or $S=1$ spin operator at $\xi$-sublattice site in  $i$-th unit cell.
The positions of seven sublattices are denoted by A, ${\rm A}^{\prime}$, B, ${\rm B}^{\prime}$, C, 
${\rm C}^{\prime}$, and D in Fig. \ref{fig1}(b). 
Note that the last term of eq. \ref{Hamiltonian} is Zeeman term.
The number of spin sites is denoted by $N$. 
Therefore, saturation magnetization value is $M_{\rm s}=SN$ where $M$ is equal to the $z$-component of the total spin $S_{\rm tot}^{z}$. 
The number of unit cells is $N/7$; we consider $N/14$ is an integer. 
Energies are measured in unit of $J$; 
we fixed $J=1$ hereafter. 
In what follows, we examine the magnetization curve in the case of $J_{\rm F}=-1$ by means of the DMRG method. 

\section{Results}
\begin{figure}
\begin{center}
\includegraphics[%
  width=1.15\linewidth,
  keepaspectratio]{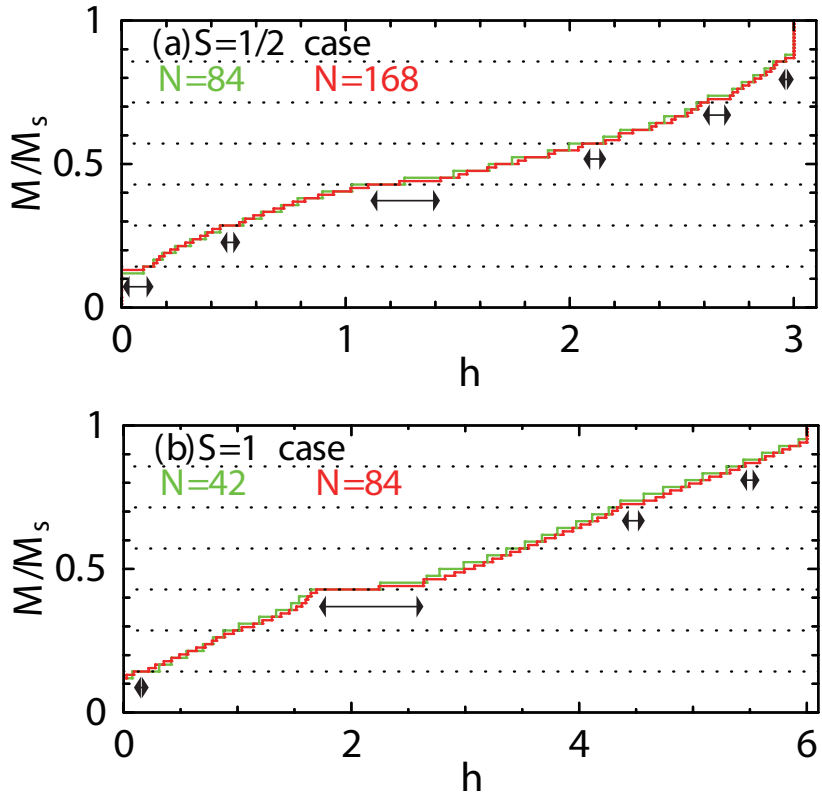}
\end{center}
\caption{(Color online) Magnetization curves of the kagome-strip lattice depicted in Fig. \ref{fig1}(b). Panels (a) and (b) are the results of $S=1/2$ and $S=1$ respectively.
Double-headed arrows indicate the regions of each magnetization plateau in the $N=168$ case for $S=1/2$ and in the $N=84$ case for $S=1$ (see also the corresponding texts).
}
\label{fig2}
\end{figure}

We show our calculation results of the magnetization curves in Fig. \ref{fig2}(a) for $S=1/2$ and in Fig. \ref{fig2}(b) for $S=1$.
In the magnetization curve of $S=1/2$, we find 6 plateaus with magnetization $M/M_{\rm s}=$ 1/7, 2/7, 3/7, 4/7, 5/7 and 6/7 where $M_{\rm s}$ means saturated magnetization 
although we should pay careful attention to the finite-size effects especially under the open-boundary condition: in finite system, 
we regard 
the region of $(\frac{1}{7}-\frac{2}{N})\leq M/M_{\rm s}\leq \frac{1}{7}$ as the 1/7 plateau, 
the region of $\frac{3}{7}\leq M/M_{\rm s}\leq (\frac{3}{7}+\frac{2}{N})$ as the 3/7 plateau,
the region of $\frac{5}{7}\leq M/M_{\rm s}\leq (\frac{5}{7}+\frac{2}{N})$ as the 5/7 plateau and
the region of $\frac{6}{7}\leq M/M_{\rm s}\leq (\frac{6}{7}+\frac{2}{N})$ as the 6/7 plateau.
In the magnetization curve of $S=1$, on the other hand, we find 4 plateaus with magnetization $M/M_{\rm s}=$ 1/7, 3/7, 5/7 and 6/7.
Note here that we regard 
the region of $\frac{3}{7}\leq M/M_{\rm s}\leq (\frac{3}{7}+\frac{1}{N})$ as the 3/7 plateau,
the region of $M/M_{\rm s}=(\frac{5}{7}+\frac{1}{N})$ as the 5/7 plateau and
the region of $M/M_{\rm s}=(\frac{6}{7}+\frac{1}{N})$ as the 6/7 plateau in finite system.
It is difficult to judge by using our present results whether or not the plateaus of $M/M_{\rm s}=2/7$ and 4/7 exists in the thermodynamic limit.
The issue of establishing the presences or absences of the 2/7 and 4/7 plateaus should be clarified in future studies.

Oshikawa-Yamanaka-Affleck theorem\cite{OYA} provides us the necessary condition for the magnetization plateaus as 
\begin{eqnarray}
Q(S-m)={\rm int.}
\end{eqnarray}
where $Q$ is the spatial periodicity of the wave function and $m$ is
the magnetization per site ($m=M/N= MS/M_{\rm s}$).
This condition tell us that the translational symmetry breaking should occur spontaneously in the magnetization plateaus of $M/M_{\rm s}=n/7$ where $n$ is even number\cite{TSB1, TSB2}.
In order to confirm the symmetry breaking, we calculate the local magnetization $\langle S_{i, \xi}^{z} \rangle$, where $\langle A \rangle$ denotes the expectation value of the physical quantity $A$ 
and $S_{i, \xi}^{z}$ is the $z$-component of ${\bf S}_{i, \xi}$.

\begin{figure}
\begin{center}
\includegraphics[%
  width=1.05\linewidth,
  keepaspectratio]{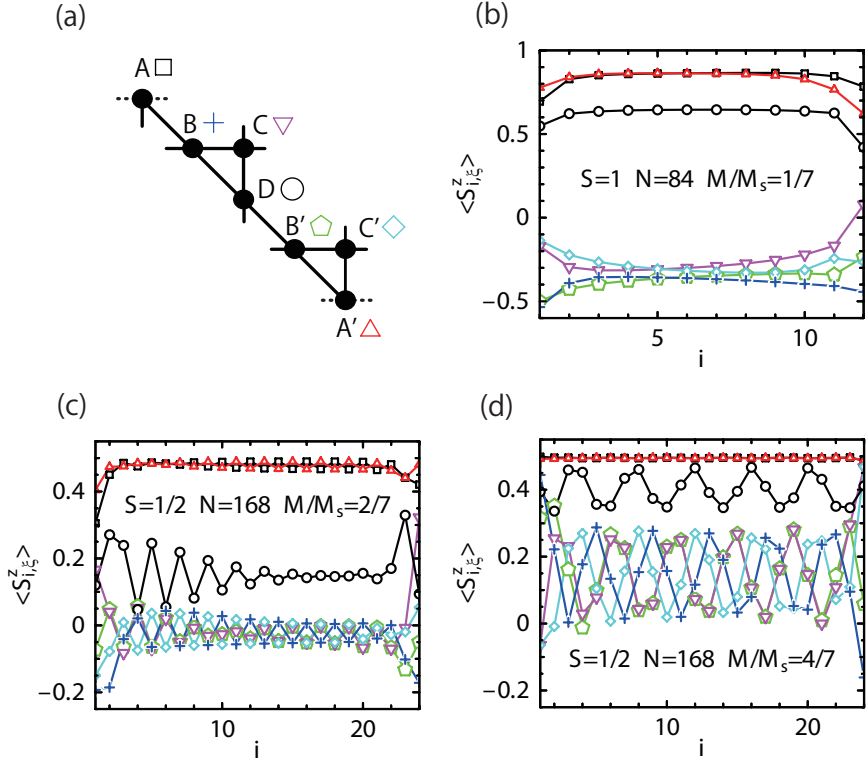}
\end{center}
\caption{(Color online) (a)The correspondence relationships between each colored symbol and each sublattice $\xi$ used in Figs. \ref{fig3}(b), \ref{fig3}(c) and \ref{fig3}(d).
Panels (b), (c) and (d) are the calculation results of the local magnetization at each sublattice $\xi$.}
\label{fig3}
\end{figure}

In Fig. \ref{fig3}(a), we first present the correspondence relationships between each colored symbol and each sublattice $\xi$ used in Figs. \ref{fig3}(b), \ref{fig3}(c) and \ref{fig3}(d).
For example of the local magnetization on the plateaus without symmetry breaking,  we show the calculation result of the local magnetization on the 1/7 plateau for the $S=1$ case in Fig. \ref{fig3}(b).
One can immediately confirm the collinear spin-configuration reflecting the present situation that there are 7 sublattices in a unit cell of the lattice depicted in Fig. \ref{fig1}(b).
%On the other hand, we confirm the spontaneous symmetry breakings in the local magnetizations such in the case of $2/7$ and $4/7$ plateaus for $S=1/2$ depicted in Fig. \ref{fig3}(c) and \ref{fig3}(d) respectively.
In Fig. 3(c) and 3(d), on the other hand, we detect the oscillation in the local magnetizations within the same sublattices in the case of 2/7 and 4/7, respectively.
The oscillation suggests that the spontaneous symmetry breakings occurs and that the wave function is degenerate. 

Finally, we discuss the relationships between our strip model and the original antiferromagnetic kagome model.
We successfully find the macroscopic jumps near the saturation field in the magnetization curves of our strip models not only in the $S=1/2$ but also in the $S=1$ cases as well as 
in the case of the $S=1/2$ and $S=1$ antiferromagnetic Heisenberg model on the original kagome lattice depicted in Fig. \ref{fig1}(a) \cite{KGM3}.
This characteristic behavior was also observed in the case of some frustrated systems\cite{jump1, jump2} and was proven in ref. 22.
%Unfortunately, on the other hand, we does not observe the one of the characteristic behaviors of the kagome antiferromagnet, magnetization ramp, in our strip models.
In our strip models, unfortunately, we does not observe clearly characteristic behavior of the magnetization ramp in the two-dimensional kagome-lattice antiferromagnet; 
Further examinations are required to capture the behavior just outside the flat region with respect to magnetization.

\section{Conclusions}
We have studied the magnetization curves of $S=1/2$ and $S=1$ Heisenberg models on the kagome strip lattice depicted in Fig. \ref{fig1}(b) by the DMRG method.
For $S=1/2$ case, we have confirmed 7 magnetic plateaus of $M/M_{s}=1/7, 2/7, 3/7, 4/7, 5/7$ and 6/7.
In the case of $M/M_{\rm s}=$2/7, 4/7, 6/7, we confirm the occurrence of the translational symmetry breaking from our numerical results of the local magnetizations.
These symmetry breakings are suggested by the Oshikawa-Yamanaka-Affleck condition.
For $S=1$ case, on the other hand, we have confirmed 4 magnetic plateaus of $M/M_{\rm s}=1/7, 3/7, 5/7$ and $6/7$.
We have also found the occurrence of the macroscopic jump in these magnetization curves near the saturation field irrespective of the spin amplitude 
as well as the $S=1/2$ and $S=1$ antiferromagnetic Heisenberg model on the original kagome lattice depicted in Fig. \ref{fig1}(a).

\begin{acknowledgements}
This work was partly supported  
by Grants-in-Aid 
(Nos. 20340096, 23340109, 23540388, and 24540348) 
from the Ministry of Education, Culture, Sports, 
Science and Technology of Japan.
DMRG calculations were carried out 
using the ALPS DMRG application\cite{ALPS}.
Part of the computations were performed 
using the facilities of 
the Supercomputer Center, Institute for Solid 
State Physics, University of Tokyo.
\end{acknowledgements}

%\pagebreak

\end{document}